\documentclass[aps,prl,twocolumn,groupedaddress]{revtex4-1}

\usepackage{graphicx}
\usepackage{amsfonts}
\usepackage{amsmath}
\usepackage{amssymb}
\usepackage{braket}

\newcommand{\state}[3]{$#1#2_{#3}$}
\newcommand{\spinup}{\ket{\uparrow}}
\newcommand{\spindown}{\ket{\downarrow}}

\begin{document}

\title{Decoherence-assisted spectroscopy of a single Mg$^+$ ion} 

\author{G.~Clos}
\email{govinda.clos@physik.uni-freiburg.de}
\affiliation{Albert-Ludwigs-Universit{\"a}t Freiburg, Physikalisches Institut, Hermann-Herder-Stra{\ss}e 3, 79104 Freiburg, Germany}
\author{M.~Enderlein}
\email[Present address: TOPTICA Photonics AG, Lochhamer Schlag 19, 82166 Gr{{\"a}}felfing, Germany]{}
\affiliation{Albert-Ludwigs-Universit{\"a}t Freiburg, Physikalisches Institut, Hermann-Herder-Stra{\ss}e 3, 79104 Freiburg, Germany}
\author{U.~Warring}
\affiliation{Albert-Ludwigs-Universit{\"a}t Freiburg, Physikalisches Institut, Hermann-Herder-Stra{\ss}e 3, 79104 Freiburg, Germany}
\author{T.~Schaetz}
\affiliation{Albert-Ludwigs-Universit{\"a}t Freiburg, Physikalisches Institut, Hermann-Herder-Stra{\ss}e 3, 79104 Freiburg, Germany}
\author{D.~Leibfried}
\affiliation{National Institute of Standards and Technology, 325 Broadway, Boulder, Colorado 80305, USA}

\date{\today}

\begin{abstract}
We describe a high-resolution spectroscopy method, in which the detection of single excitation events is enhanced by a complete loss of coherence of a superposition of two ground states. Thereby, transitions of a single isolated atom nearly at rest are recorded efficiently with high signal-to-noise ratios. Spectra display symmetric line shapes without stray-light background from spectroscopy probes. We employ this method on a $^{25}$Mg$^+$ ion to measure one, two, and three-photon transition frequencies from the \state{3}{S}{} ground state to the \state{3}{P}{}, \state{3}{D}{}, and \state{4}{P}{} excited states, respectively. Our results are relevant for astrophysics and searches for drifts of fundamental constants. Furthermore, the method can be extended to other transitions, isotopes, and species. The currently achieved fractional frequency uncertainty of $5 \times 10^{-9}$ is not limited by the method.
\end{abstract}
\pacs{32.30.Jc, 37.10.Ty, 37.10.Rs, 37.25.+k}
\maketitle
%
Quantum systems that are well isolated from their environments, e.g., tailored solid-state systems, photons, and trapped atoms, offer a high level of control~\cite{ladd_quantum_2010}. Over the past decades, several experimental methods have been devised for quantum control of single trapped ions~\cite{brown_geonium_1986,wineland_nobel_2013,leibfried_quantum_2003}. Developments are driven by the urge to make more accurate and precise clocks~\cite{diddams_optical_2001,schmidt_spectroscopy_2005} as well as to address questions in different fields of research, e.g., properties of highly charged ions~\cite{gillaspy_highly_2001,derevianko_highly_2012}, ion-neutral collisions~\cite{cote_ultracold_2000,idziaszek_controlled_2007,bodo_ultra-cold_2008,idziaszek_quantum_2009}, molecular physics~\cite{carr_cold_2009,kahra_molecular_2012,loh_precision_2013}, and tests of fundamental physics~\cite{hanneke_new_2008,pohl_size_2010,atrap_collaboration_one_2013,mooser_resolution_2013,disciacca_resolving_2013}. High-resolution spectroscopy measurements~\cite{drullinger_high-resolution_1980,rosenband_frequency_2008,wolf_frequency_2008,batteiger_precision_2009,wan_precision_2013} are of particular interest for studying spatial and temporal fine structure variations of the universe~\cite{webb_search_1999,dzuba_space-time_1999,webb_indications_2011}. In such experiments, complex atomic and molecular structures need to be probed by single- or multi-photon transitions in isotopically pure samples revealing undisturbed transition line shapes. Weak transitions in trapped ions can be measured with various methods~\cite{schmidt_spectroscopy_2005,wineland_nobel_2013}, and techniques based on the detection of momentum kicks altering the occupation of motional states from few absorbed photons have been developed that are applicable to strong electric dipole transitions as well~\cite{hempel_entanglement-enhanced_2013,wan_precision_2013}. In this Letter, we experimentally study single- and multi-photon transitions in a single, laser cooled $^{25}$Mg$^+$ ion that can be near-perfectly isolated from its environment.
We detect the decoherence  of a superposition of two electronic ground states due to single scattering events and determine transition frequencies which are relevant for astrophysics and searches for variations of fundamental constants~\cite{berengut_laboratory_2010,destree_detection_2010} with a fractional uncertainty of $5\times10^{-9}$.

For a simplified description of the method, consider an atom with three states. Two of these states, labeled $\spinup$ and $\spindown$, which are long-lived and allow for coherent control, are used to study transitions to a third, excited state $\ket{e}$.
After preparation in $\spinup$, a $\pi/2$ pulse on the $\spinup \rightarrow \spindown$ transition creates a superposition state $\ket{\psi} \equiv \frac{1}{\sqrt{2}}\left( \spinup + \spindown\right)$. A spectroscopy pulse probes the couplings $\spinup \rightarrow \ket{e}$ and $\spindown \rightarrow \ket{e}$ during a delay period $\tau$. To decouple the superposition state from sources of detrimental decoherence, a spin-echo pulse is applied after $\tau/2$. A second $\pi/2$ pulse completes the sequence and the final state is analyzed.
Disregarding the influence of the spectroscopy pulse, the total sequence coherently transfers $\spinup$ to $\spinup$.
However, absorption of a probe photon and the subsequent spontaneous emission project the system into $\spinup$ or $\spindown$, i.e., the original phase information of $\ket{\psi}$ is destroyed. The remaining sequence creates a new superposition state of $\spinup$ and $\spindown$, and analyzing the final state, the detection probability of single excitations is 1/2.
Hence, decoherence constitutes the spectroscopic signal and, therefore, the method probes only excitation strengths and is insensitive to branching ratios of the spontaneous decay, making it a versatile method to study a variety of transitions.

%
\begin{figure}[htb]
	\centering
		\includegraphics[width=8.6cm]{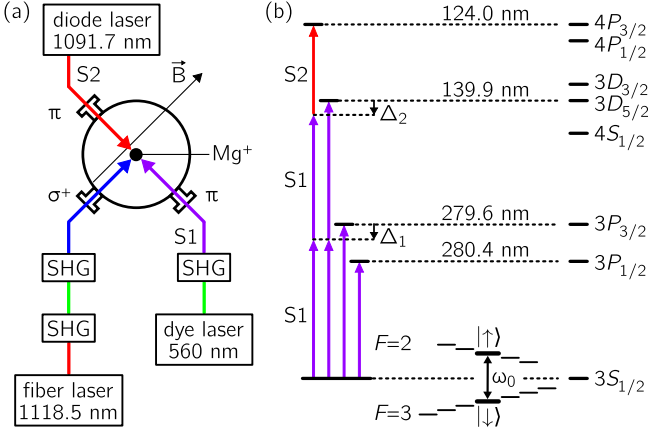}
	\caption{(color online)
Schematic of the experimental setup and diagram of relevant states of $^{25}$Mg$^+$ (nuclear spin 5/2).
(a) The vacuum chamber and laser setups including SHG. The fiber laser provides Doppler cooling, state preparation and detection. The dye (S1) and diode (S2) laser serve as dedicated spectroscopy lasers.
(b) Energy level diagram (not to scale) including fine structure levels up to \state{4}{P}{3/2}. For the ground state, the hyperfine and Zeeman splitting are shown. Arrows between levels indicate the energies utilized in our one, two, and three-photon spectroscopy transitions, and relevant detunings to intermediate states are labeled $\Delta_1$ and $\Delta_2$.
}
	\label{fig:setup}
\end{figure}
In our experimental demonstration of the method, we trap a single $^{25}$Mg$^+$ ion in a linear Paul trap~\cite{schaetz_towards_2007}. A schematic of the setup is shown in Fig.~\ref{fig:setup}(a), while the relevant energy levels are depicted in Fig.~\ref{fig:setup}(b). The beam of a fiber laser, frequency-quadrupled using two second harmonics generation (SHG) stages~\cite{friedenauer_high_2006}, is tuned $\Gamma_{\rm nat}/2$ below the \state{3}{S}{1/2} to \state{3}{P}{3/2} transition (natural line width $\Gamma_{\rm nat}/(2\pi)= 41.8(4)\;\rm MHz$~\cite{ansbacher_precision_1989}) and aligned with a magnetic quantization field $\vert \vec{B} \vert \simeq0.58\rm\;mT$. The light is $\sigma^+$-polarized and provides Doppler cooling to $\simeq 1\;\rm mK$ and optical pumping into the state \state{3}{S}{1/2}~$\ket{F{=}3, m_F{=}3}$, where $F$ and $m_F$ denote the total angular momentum quantum numbers of the valence electron. This state provides efficient, state sensitive detection via closed cycling transition  to \state{3}{P}{3/2}~$\ket{4,4}$ to discriminate between the ground state manifolds. We observe a count rate of $100\;\rm ms^{-1}$ on average for all $F{=}3$ states and $2\;\rm ms^{-1}$ for the $F{=}2$ states. As parts of the superposition state $\ket{\psi}$ we choose the low field clock states $\spinup\equiv$ \state{3}{S}{1/2}~$\ket{2,0}$ and $\spindown\equiv$ \state{3}{S}{1/2}~$\ket{3,0}$ which are separated by $\omega_0/(2\pi)\simeq 1.789\rm\;GHz$. These states feature a low sensitivity to magnetic field fluctuations and long coherence times. We employ microwave pulses to transfer population from \state{3}{S}{1/2}~$\ket{3,3}$ to $\spinup$ with near-unity fidelity, and to coherently control $\spinup$ and $\spindown$.
 For spectroscopy we use two laser systems: a frequency-doubled dye laser (S1) at wavelength $\lambda\simeq280\rm\;nm$ with a beam waist radius ($1/e^2$ radius of intensity) of  $w=25(3)\;\mu$m, and a diode laser (S2, $\lambda\simeq1092\rm\;nm$, $w=190(20)\;\mu$m). The S1 and S2 laser beams enter the chamber from opposite directions, perpendicular to $\vec{B}$ with linear polarization to induce $\pi$ transitions.
We determine the spectroscopy laser wavelengths with a wavelength meter (HighFinesse WS Ultimate/2), which is referenced to the R(53)28-3 
line in $^{127}$I$_2$ and the D$_2$ line in $^{87}$Rb via Doppler-free spectroscopy.

Since $^{25}$Mg$^+$ has more than two hyperfine levels in the ground state, the final state is not restricted to superpositions of $\spinup$ and $\spindown$, and the excited state can decay into other hyperfine ground states, analogous to electron shelving~\cite{leibfried_quantum_2003}. In that case, the subsequent microwave pulses are off resonant leaving the state unchanged. These other decay channels also alter the detection outcome when recording the spectrum via state-dependent fluorescence. However, comparing to spectra recorded by optical pumping and electron shelving techniques, our method reveals additional features, e.g., excitations from $\spinup$ to \state{3}{P}{3/2}~$\ket{F{=}1, m_F{=}0}$, $\spindown$ to \state{3}{P}{3/2}~$\ket{F{=}4, m_F{=}0}$ (see below), as well as cycling transitions. Furthermore, it facilitates the detection of transitions with unfavorable branching ratios of the spontaneous decay, and the interlaced $\pi$ pulse enables extended probe periods to detect weak transitions.

\begin{figure}[tb]
	\centering
       		\includegraphics[width=8.4cm]{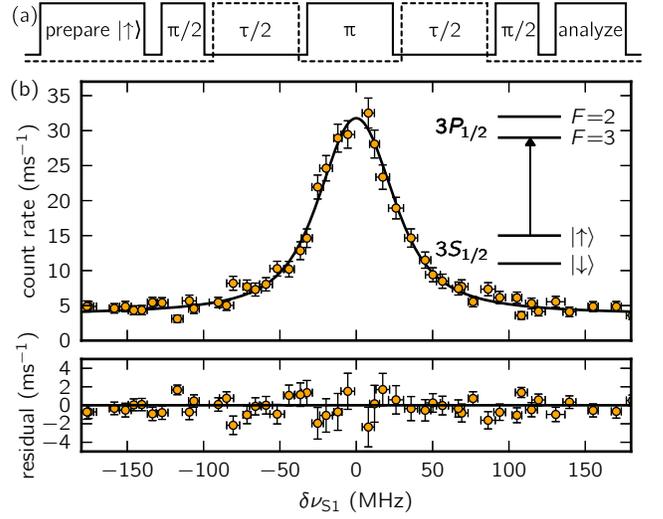}
		\includegraphics[width=8.6cm]{figure2-2.eps}
        \caption{(color online)
Pulse sequence and exemplary absorption spectrum.
(a) Schematic of the pulse sequence used for spectroscopy (details see text). The dashed line indicates the spectroscopy laser sequence. The final state is analyzed via state-dependent fluorescence during a period of $20\;\mu \rm s$. Hence, the spectrum is recorded without stray-light background from the spectroscopy lasers. 
(b) Absorption spectrum of the $\spinup$  to \state{3}{P}{1/2}~$\ket{3,0}$ transition.
The error bars in both dimensions indicate the statistical uncertainty. A Voigt fit to the data (solid line) is consistent with a fixed Lorentzian width $\Gamma_{\rm nat}$ and an extracted Gaussian width of $\Gamma_G/(2\pi)=30(7)\;\rm MHz$. The fit residuals show no additional systematics.
}
	\label{fig:p12}
\end{figure}
%
We first implement the method on the \state{3}{S}{1/2} to \state{3}{P}{1/2} transition, using the pulse sequence outlined in Fig.~\ref{fig:p12}(a). The spectroscopy laser S1 is applied for $\tau = 0.2\;\rm ms$ with an intensity $I_{\rm S1} = 1.5(3)\;\rm W/m^2$. This corresponds to an on-resonance saturation parameter $s_0 = I/I_{\rm sat}=6(1)\times 10^{-4}$ with saturation intensity $I_{\rm sat}\simeq 2500\;\rm W/m^2$, and is equivalent to on average two scattering events per experiment.
 We repeat the sequence and gradually increase the frequency of S1 to span $2.5\;\rm GHz$ resulting in an absorption spectrum with two resonances. These are separated by $\omega_0$ and the excited-state hyperfine structure.
 Each collected data point represents the average of 2000 experiments, acquired in about $5\;\rm s$ per point.  Fig.~\ref{fig:p12}(b) only covers a fraction of the full spectrum to emphasize the undistorted, symmetric line shape centered at the $\spinup$  to \state{3}{P}{1/2}~$\ket{3,0}$ resonance.
 A best fit is shown as a solid line and yields no significant deviation from a Voigt profile with fixed natural line width $\Gamma_{\rm nat}/(2\pi)=41.3(3)\;\rm MHz$~\cite{ansbacher_precision_1989}. 
We find a Gaussian contribution of $\Gamma_G/(2\pi)=30(7)\;\rm MHz$ while the calculated Doppler limit amounts to $\Gamma_{\rm Doppler}/(2\pi) \simeq5\;\rm MHz$ and the broadening due to residual micromotion is estimated to be less than $0.5\;\%$ of the natural line width. Based on a beat note measurement of S1 with the Doppler cooling laser (at $280\;\rm nm$), we attribute the main part of the Gaussian contribution to the line width of S1. Since this broadening effect preserves symmetric line shapes, the center of the resonance is still determined with high precision.

%
\begin{figure*}
\centering
        \includegraphics[width=17.8cm]{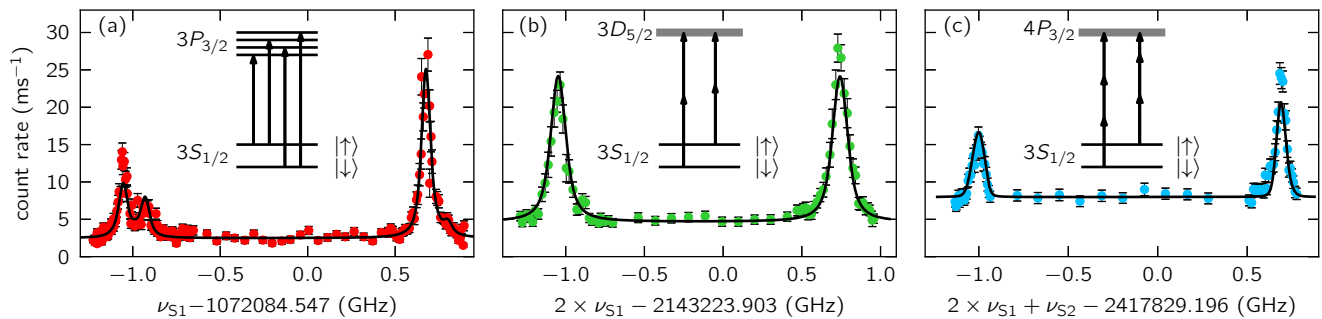}
        \caption{(color online)
Recorded absorption spectra for one, two, and three-photon transitions including model fits to the data shown as solid lines (details see text). The error bars include the statistical uncertainty both in count rate and in frequency.
For each transition, we depict the corresponding energy level diagram (not to scale).
(a) One-photon transition \state{3}{S}{1/2} to \state{3}{P}{3/2} resolving the excited state hyperfine splitting ($\tau = 0.2\;\rm ms$, $I_{\rm S1} = 0.5(1)\;\rm W/m^2$). Here, one can see two transitions that cannot be detected by electron shelving spectroscopy.
(b) In the two-photon \state{3}{S}{1/2} to \state{3}{D}{5/2} absorption spectrum the underlying substructure lies within the natural line width ($\tau = 0.2\;\rm ms$, $I_{\rm S1} = 1.8(4)\;\rm MW/m^2$). 
(c) Three-photon transition connecting \state{3}{S}{1/2} to \state{4}{P}{3/2}, where the substructure shows as slight asymmetries in the spectra ($\tau = 2.0\;\rm ms$, $I_{\rm S1} = 2.6(5)\;\rm MW/m^2$, $I_{\rm S2} = 160(30)\;\rm kW/m^2$).
}
\label{fig:123photon}
\end{figure*}
In total, we study four different transitions to excited electronic states in detail: two one-photon transitions (\state{3}{S}{1/2} to \state{3}{P}{1/2} and \state{3}{P}{3/2}), one two-photon transition (\state{3}{S}{1/2} to \state{3}{D}{5/2}), and one three-photon transition (\state{3}{S}{1/2} to \state{4}{P}{3/2}). These transitions, as well as the relevant detunings $\Delta_1$ and $\Delta_2$ are sketched in Fig.~\ref{fig:setup}(b). We record 28 one-photon transition spectra with different laser intensities $I_{\rm S1} = 0.05$--$3\;\rm W/m^2$ and probe durations $\tau = 0.2$--$3.2 \;\rm ms$. Figure~\ref{fig:123photon}(a) shows an absorption spectrum of the  \state{3}{S}{1/2} to \state{3}{P}{3/2} transition that resolves the substructure resulting from hyperfine interactions.
The two-photon transition to \state{3}{D}{5/2} ($\Gamma_{\rm nat}/(2\pi)\simeq77.7\;\rm MHz$~\cite{NIST_lifetimes}) is probed with intensities $I_{\rm S1} = 0.6$--$2.2\;\rm MW/m^2$ and probe durations $\tau = 0.2$--$0.6 \;\rm ms$ leading to 24 different spectra centered around a detuning $\Delta_1/(2\pi) \simeq -500\;\rm GHz$ of S1 from \state{3}{P}{3/2}; one representative
 is shown in Fig.~\ref{fig:123photon}(b).
In the three-photon absorption spectra to \state{4}{P}{3/2} ($\Gamma_{\rm nat}/(2\pi)\simeq8.5\;\rm MHz$~\cite{NIST_lifetimes}), we vary the frequency of S2 in discrete steps to obtain a spectrum, while the frequency of S1 is locked via Doppler-free spectroscopy at $\Delta_1/(2\pi) \simeq -500\;\rm GHz$ from \state{3}{P}{3/2} and $\Delta_2/(2\pi) \simeq 70\;\rm MHz$ from \state{3}{D}{5/2}, see Fig.~\ref{fig:123photon}(c). Due to the hyperfine splitting of the ground state, the detuning $\Delta_2/(2\pi)$ from the line centroid amounts to roughly $\pm 900\;\rm MHz$ from two-photon resonance (cf.\ Figs.~\ref{fig:setup}(b) and \ref{fig:123photon}(b)). We record 28 spectra with varying laser intensities $I_{\rm S1} = 1.4$--$3.5\;\rm MW/m^2$ and $I_{\rm S2} = 30$--$130\;\rm kW/m^2$, and probe durations $\tau = 0.2$--$2.0 \;\rm ms$.

To extract the line centroids from the data, each absorption spectrum is analyzed by a best fit considering the substructure as a sum of Voigt profiles. To reduce the number of free fit parameters, relative transition strengths and sublevel splittings due to hyperfine and magnetic field interactions as well as the effect of population transfer between $\spinup$ and all other hyperfine ground states due to spontaneous emission are set to their calculated values~\cite{metcalf}. The calculations include magnetic-dipole and electric-quadrupole hyperfine constants~\cite{itano_precision_1981,sur_comparative_2004,safronova_relativistic_1998}, and natural line widths. We calibrate the magnetic field strength $\vert \vec{B} \vert = 0.5848(3)\;\rm mT$ using measurements of \state{3}{S}{1/2}~$\ket{3,3}$ to $\ket{2,2}$ microwave transitions with a magnetic field sensitivity of $\simeq 24 \;\rm MHz/mT$.  We neglect transitions with a calculated relative intensity of less than $10^{-3}$, leaving four free fit parameters: fluorescence offset, signal intensity, Gaussian width $\Gamma_G$, and centroid frequency $\nu_c$. The results are shown as solid lines in Fig.~\ref{fig:123photon}.
Within our measurement resolution, we find no systematic shifts of the line centroids of the one and two-photon transitions for our range of intensities and pulse durations. However, probing the three-photon transition, we observe a systematic shift related to the intensity of S2 of up to $10\;\rm MHz$. We account for this effect by including ac-Stark shifts from off-resonant coupling of S2 to the \state{3}{D}{5/2} and \state{4}{P}{3/2} states into our fit model.
 To this end, for the three-photon transition, we introduce the intensity of S2 as an additional free parameter in the fit routine. 
We compare the fitted intensities with the results of beam waist and power measurements and find that the observed shift is consistent with being entirely due to this effect.
  
\begin{figure}[tb]
	\centering
		\includegraphics[width=8.6cm]{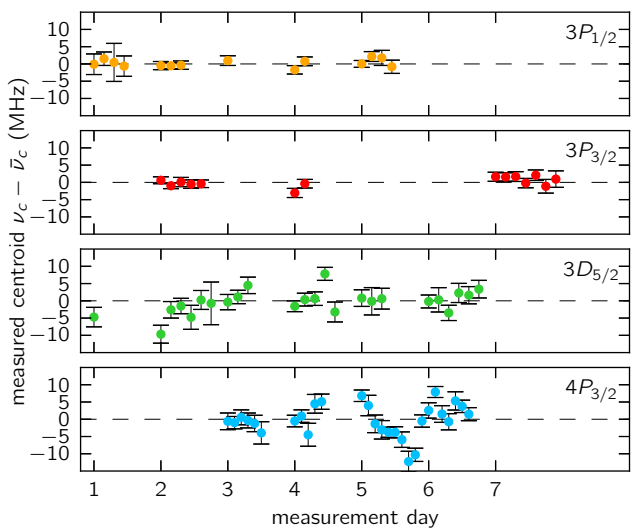}
	\caption{(color online)
Line centroids for  the  four different transitions measured during seven days. The values are plotted relative to their mean values (dashed lines, cf.\ Table~\ref{tab:results}) and error bars indicate the statistical uncertainty of the values extracted from individual model fits. The measurements of the transition frequencies to the \state{3}{P}{} states are in excellent agreement with Ref.~\cite{batteiger_precision_2009}. The different measurements illustrate the reproducibility of the center frequency determinations beyond their estimated systematic uncertainty. The results for the \state{3}{D}{} and \state{4}{P}{} measurements show a statistical spread that is significantly smaller than this uncertainty.
}
	\label{fig:stability}
\end{figure}

In Fig.~\ref{fig:stability} we show the deviation of each fitted $\nu_c$ from the mean centroid $\bar{\nu}_c$ of all analyzed spectra, with the data ordered according to the measurement day. Each data point represents a spectrum similar to those shown in Fig.~\ref{fig:123photon} and the error bars indicate the statistical uncertainty from the fit result, while the corresponding $\bar{\nu}_c$ are listed in Table~\ref{tab:results}.
 The main systematics dominating the estimated uncertainty of the transition frequencies stems from the frequency measurement. We conservatively estimate this to be 
$2\;\rm MHz$ near $560 \;\rm nm$ and $10\;\rm  MHz$ near $1092\;\rm nm$. 
 We calibrated the wavelength meter with two reference frequencies, known to an uncertainty of $1.5\rm\; MHz$ for the $^{127}\rm I_2$ line~\cite{iodine_spec} and $0.2\;\rm MHz$ for the $^{87}\rm Rb$ line~\cite{ye_hyperfine_1996,bize_high-accuracy_1999}. 
 When probing the three-photon transition, an additional systematic uncertainty of $4\;\rm MHz$ arises from the Doppler-free spectroscopy used for locking S1. Additional effects, e.g., correction of ac Stark shifts, magnetic field fluctuations, spectral widths of the spectroscopy lasers, and deviations of relative intensities, each contribute less than $0.5 \;\rm MHz$ to the final uncertainties.
\setlength{\tabcolsep}{4pt}
\begin{table}
\renewcommand{\arraystretch}{1.5}
\begin{tabular}{lll}
\multicolumn{3}{c}{Transition frequencies $\bar{\nu}_c$ in $^{25}$Mg$^+$ from \state{3}{S}{1/2} in THz}\\
\toprule
&\state{3}{P}{1/2}&\state{3}{P}{3/2}\\
\colrule
This work&1\,069.339\,957\,(5)&1\,072.084\,547\,(5)\\
Batteiger et al.~\cite{batteiger_precision_2009}& 1\,069.339\,96\,(2)&1\,072.084\,56\,(2)\\
\toprule
& \state{3}{D}{5/2}&\state{4}{P}{3/2}\\
\colrule
This work&2\,143.223\,903\,(7)&2\,417.829\,196\,(12)\\
Martin et al.~\cite{martin_energy_1980}&2\,143.222\,0\,(15)&2\,417.826\,8\,(15)\\
Goorvitch et al.~\cite{goorvitch_vacuum-ultraviolet_1970}&2\,143.227\,7\,(18)&2\,417.805\,(10)\\
\botrule
\end{tabular}
\caption{
Transition frequencies (including statistical and systematic uncertainties) determined in our work in comparison to representative literature values. Our values for the \state{3}{P}{} states are in agreement with previous results~\cite{batteiger_precision_2009}. For the \state{3}{D}{5/2} and \state{4}{P}{3/2} state we have agreement within two standard deviations with a more than two orders of magnitude improvement in fractional frequency uncertainty compared to Refs.~\cite{martin_energy_1980,goorvitch_vacuum-ultraviolet_1970}.}
\label{tab:results}
\end{table}
%

 
In Table~\ref{tab:results} we compare our results to literature values. The one-photon transition frequencies are in good agreement with the values from Ref.~\cite{batteiger_precision_2009}. We have conducted the first isotopically pure measurements of transition frequencies to the \state{3}{D}{5/2} and \state{4}{P}{3/2} levels, and improved the fractional frequency uncertainty compared to previously calculated~\cite{martin_energy_1980} and experimental~\cite{goorvitch_vacuum-ultraviolet_1970} values by more than two orders of magnitude.

We record high-SNR and symmetric line shapes of one, two, and three-photon transitions, allowing for the determination of transition frequencies with a fractional frequency uncertainty of $5 \times 10^{-9}$. This uncertainty is limited by our wavelength measurement and can be substantially improved by using a frequency comb. In addition, the resolution can be enhanced by use of spectroscopy lasers with smaller spectral widths and by cooling the ion to the motional ground state.
With the demonstrated sensitivity and the multitude of accessible transitions enabled by utilizing a superposition state, the method may facilitate the determination of transition strengths, natural line widths, and hyperfine constants. Furthermore, it can be extended to other transitions, isotopes, species, and even other quantum systems. In particular, the signal induced by decoherence is insensitive to branching ratios of the spontaneous decay. In combination with a logic ion~\cite{wineland_nobel_2013}, and incorporating the motional degrees of freedom into our method, it may also be applicable to species without cooling and detection transitions as well as molecular ions.
 We anticipate that this spectroscopy method is only one of many future techniques that take advantage of decoherence effects that are typically thought of as detrimental when controlling quantum systems.
%

\begin{acknowledgments}
This work was supported by DFG (SCHA 973), and
EU (PICC, Grant No.~249958). 
We gratefully acknowledge the technical support from the company HighFinesse,
who provided us with the wavelength meter. We thank M.\ Bujak for installing the diode laser system, and J.\ Denter for
maintaining the dye lasers.
\end{acknowledgments}
%

%

\end{document}